    \documentclass[]{aa}
    \usepackage{psfig}
    \usepackage{graphicx}
    \usepackage{natbib}

    \def\iiii{}			
    \def\iii{}			
    \def\ii{}			

    \def\ergsec{\hbox{erg s$^{-1}$ }}
    \def\ergcm{\hbox{erg cm$^{-2}$ s$^{-1}$ }}
    \def\erga{\hbox{erg cm$^{-2}$ s$^{-1}$ \AA$^{-1}$ }}

    \def\Msun{$M_{\rm \odot}$}

    \def\mdot{\dot M}
    \def\phiorb{\ifmmode\phi_{\rm orb}\else$\phi_{\rm orb}$\fi}
    
    \def\it{\sl}
    \def\degs{\ifmmode ^{\circ}\else$^{\circ}$\fi}
    \def\amin{\ifmmode ^{\prime}\else$^{\prime}$\fi}
    \def\asec{\ifmmode ^{\prime\prime}\else$^{\prime\prime}$\fi}

    \def\ob{V1309~Ori}
    \def\ste{$^{\star}$}

\begin{document}


     \title{XMM-Newton observation of the long-period polar \ob : The case for 
            pure blobby accretion\thanks{Based on observations 
            obtained with XMM-Newton, an ESA
     science mission with instruments and contributions directly funded 
     by ESA member states and NASA.}}

       \author{R. Schwarz\inst{1,2} \and
               K. Reinsch\inst{2} \and
               K. Beuermann\inst{2}
               \and V. Burwitz\inst{3}
             }

      \authorrunning{R.~Schwarz et al.}
  \titlerunning{XMM-Newton observations of the long period polar V1309 Ori}
   \offprints{R. Schwarz, rschwarz@aip.de}
 
  \institute{
   Astrophysikalisches Institut
          Potsdam, An der Sternwarte 16, D--14482 Potsdam, Germany
        \and 
Universit\"atssternwarte G\"ottingen, Geismarlandstra\ss e 11, 
        D--37083 G\"ottingen, Germany 
        \and Max-Planck-Institut f\"{u}r Extraterrestrische Physik,
           Giessenbachstra\ss e, D--85740 Garching, Germany
      } 

   \date{Received ; accepted } 

\abstract{ 
Using XMM-Newton we have obtained the first continuous X-ray
observation covering a complete orbit of the longest period
polar, \ob{}. The X-ray light curve is dominated by a short, bright
phase interval with EPIC pn count rates  reaching up to 15
cts~s$^{-1}$ per 30 sec resolution bin. The bright phase
emission is well described by a  single blackbody 
component with $kT_{\rm bb} = (45 \pm 3)$ eV. The absence
of a bremsstrahlung component at photon energies above 1 keV yields a
flux ratio $F_{\rm bb}/F_{\rm br} \geq 6700$. This represents
the most extreme case of a soft X-ray excess yet observed in an
AM Herculis star. The bright, soft X-ray emission is subdivided
into a series of individual flare events supporting the
hypothesis that the soft X-ray excess in \ob\ is caused by accretion
of dense blobs  carrying the energy into sub-photospheric layers.
On average, the flares have rise and fall times of 10 sec. 
In addition to the bright phase emission,
a faint, hard X-ray component is visible throughout the
binary orbit with an almost constant count rate of 0.01
cts~s$^{-1}$. Spectral modelling indicates that this emission
originates from a complex multi-temperature plasma. At least
three components of an optically thin plasma with temperatures
$kT= 0.065$, 0.7, and  2.9 keV are required to fit the observed flux
distribution. 
The faint phase emission is occulted during the optical eclipse.
Eclipse ingress lasts about 15--20 min and is 
substantially prolonged beyond nominal ingress of the white
dwarf. This and the comparatively low plasma temperature provide
strong evidence that the faint-phase 
emission is not thermal bremsstrahlung from a post-shock accretion 
column above the white dwarf.  
A large fraction of the faint-phase emission is ascribed to the spectral
component with the lowest temperature and could be explained by scattering 
of photons from the blackbody component in the infalling material above 
the accretion region. The remaining hard X-ray flux could be produced in
the coupling region, so far unseen in other AM Herculis systems.
}

\maketitle

\section{Introduction}
{\ii The flux distribution of the high-energy emission from}
the impact region{\ii s} of strongly magnetic cataclysmic variables
(AM Herculis stars or polars) 
{\ii provides important clues on the physics of accretion onto {\iii
magnetic } white
dwarfs.}  The standard model {\ii for a magnetically collimated
accretion column} \citep{Lamb79,King79} involv{\ii es} a shock {\ii
standing} above the white dwarf {\ii surface in which the plasma is
decelerated and heated to $\sim$\,10$^8$\,K. This model} predicts
parity between the hard X-ray flux from the post-shock flow and the
reprocessed component from the irradiated white dwarf surface emitted
as a $\sim 10^{5}$~K blackbody in the soft X-ray band. Early X-ray
observation{\ii s} \citep[e.g. {\ii of} AM Her;][]{Rothschild81}
revealed that such an energy balance is strongly violated with the
soft X-ray component exceeding the shock emission by a factor of 10 or
more. An additional strong hint against this simple reprocession model
is given by the missing correlation {\ii between} the short-term
variability in the soft and hard X-ray bands
\citep{Stella86,Watson87}.  Theory challenged by this 'soft X-ray
puzzle' suggested that {\ii the} excess emission may result from {\ii
sub-photospheric energy release of} dense {\ii filaments} of material
{\ii leading to a complete thermalisation of the} accretion luminosity
{\ii and to blackbody-like emission} in the soft X-ray band
\citep{Kuijpers82}.  A detailed description of the conditions for this
'blobby' accretion scenario by \cite{Frank88} showed that densities
larger $10^{-7}$ g cm$^{-3}$ are required to reach sufficient optical
depths and are the natural consequence of stream segregation at large
coupling radii and subsequent compression in the magnetic funnel.
While blobby accretion is widely accepted as the origin of {\ii the}
soft X-ray excess, other concurrent processes may affect the observed
energy distribution in polars.  These include re-emission of hard
X-ray photons in the UV \citep{Heise88,Gaensicke95}, or a deficiency
of hard X-ray {\ii emission} due to {\ii the} dominance of cyclotron
cooling over thermal bremsstrahlung in high-field, low-density
environment{\ii s} \citep{Woelk96}.  The physics of the blob impact
itself and its effects on the white dwarf atmosphere is largely
unexplored. The {\ii best} approach to the problem {\ii so far} is
restricted to the treatment of a stationary heatflow problem
\citep{Litchfield89}.

Observationally, the soft X-ray excess has been firmly established for
a large number of polars using ROSAT PSPC observations
\citep{Beuermann95,Ramsay94}, where also a strong correlation with the
magnetic field strength {\ii in the accretion region} was found.
Recently, the status of the energy balance has been critically
reconsidered by \cite{Ramsay04} on the basis of {\ii a} re-analys{\ii
is of} ROSAT PSPC data and a snap-shot survey of 22 polars {\ii
observed} with XMM-Newton in high state{\ii s of accretion}.  They
question the existence of a soft X-ray excess for the majority of
systems and find that high instantaneous accretion rate{\ii s} lead to
high soft-to-hard {\ii X-ray flux} ratio{\ii s}.  This is in line with
the behaviour of polars observed at very low accretion rates
\citep{Ramsay04b} whose spectra do not show any distinct soft
component.

The eclipsing polar \ob\ is one of the most peculiar AM Herculis
stars.  With an orbital period of $\sim 8$~hrs, almost twice as long
as any other system, it is a key object for understanding the
evolution and synchronisation of magnetic CVs \citep{Frank95}.  {\ii
An}other puzzling aspect concerns the dominance of {\ii emission from
the} accretion stream at infra-red, optical and UV bands
\citep{Shafter95} 
cyclotron radiation from the accretion shock {\ii provides} the main
contribution.
Diagnostics of the accretion process{\ii es in \ob\ } are therefore
restricted to X-ray observations.
The most {\ii longest pointing with} ROSAT \citep{Walter95} revealed
{\ii that} \ob\ {\ii is} a strongly flaring X-ray source with a very
soft spectrum.  In this paper we present the first continuous, whole
orbit X-ray observation of \ob . 

\section{Observations and reduction}
\begin{figure}
\includegraphics[width=1.0\columnwidth,clip]{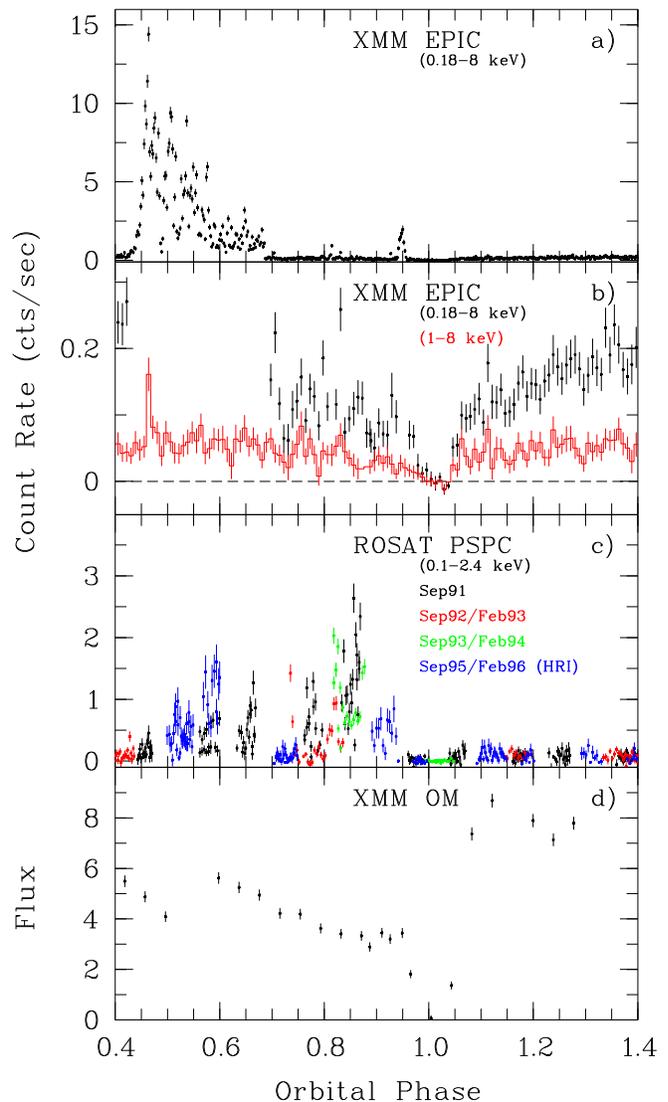}
\caption{a) Combined EPIC pn/ MOS X-ray light curves of \ob\ observed
with XMM-Newton {\ii in} March 2001.  b) Same as a) but enlarged in
count rate space to emphasise the {\ii emission components during the}
faint phase and {\ii in the} 1--8 keV hard X-ray {\ii band} (red
line).
c) 0.1--2.4 keV X-ray light curve of \ob\ obtained with the ROSAT
satellite at various epochs.  d) UV light curve from the optical
monitor {\ii onboard XMM-Newton} taken with the UVW2 filter.  {\ii
Fluxes are} given in units of $10^{-15}$\erga .  }\label{f:lcall}
\end{figure}

\begin{figure*}[t]
\includegraphics[clip=,angle=-90,width=1.99\columnwidth]{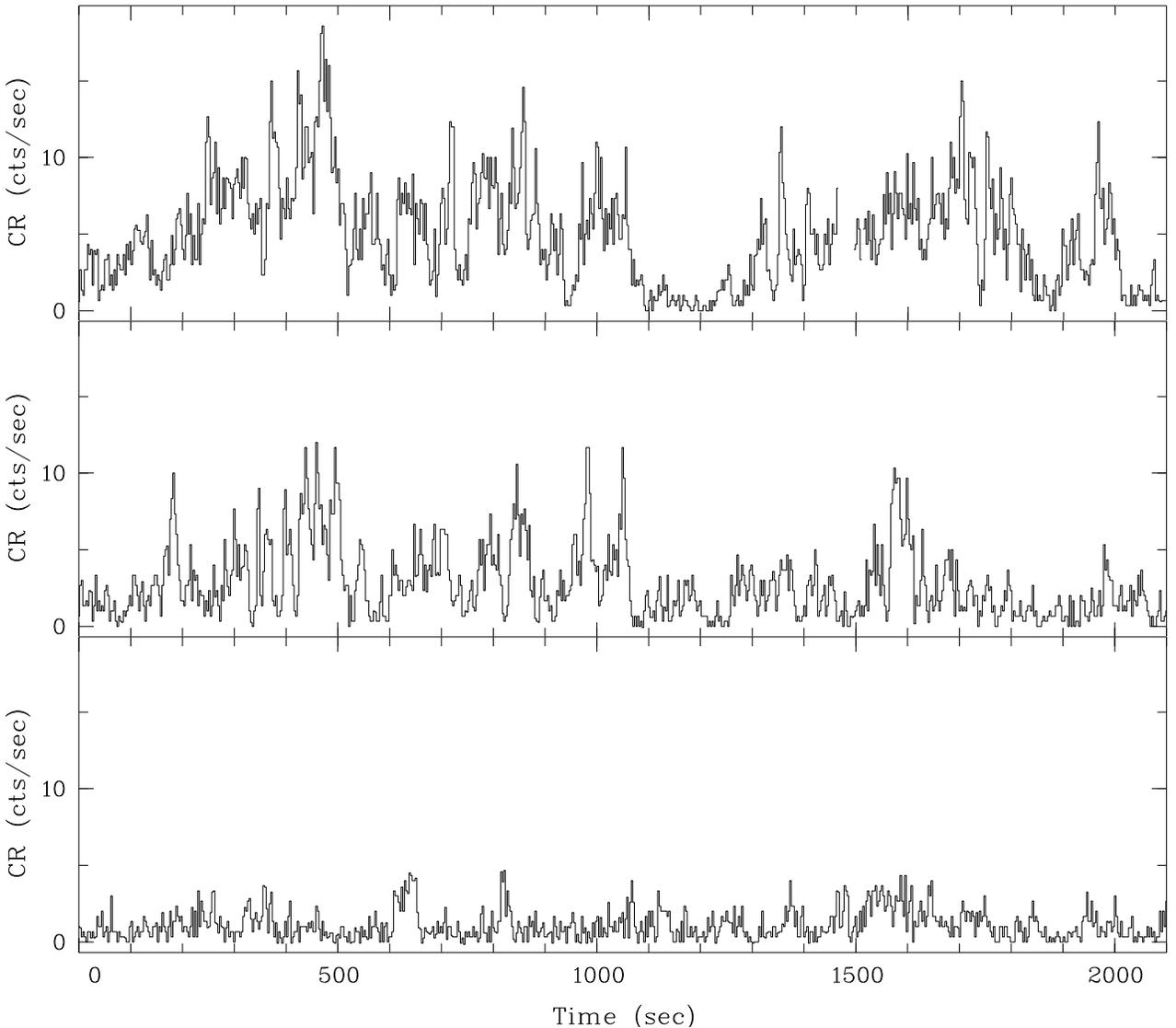}
\caption{
{\ii 0.18--8 keV EPIC pn count rates during three subsequent bright-phase 
intervals (\phiorb = 0.45--0.66) of \ob\  binned at a time resolution of 3 s.}
}\label{f:flare}
\end{figure*}

\ob\ was target of a pointed XMM-Newton observation during revolution
\#233 starting on March 18, 2001.  Our analysis focuses on data from
the European Photon Counting Cameras (EPIC) which were all operated
with a thin filter.  With a total exposure time of 29.1 ksec the {\ii
observations with the two} MOS {\ii detectors} covered just one
complete cycle of \ob , while the shorter pn exposure (on-time
$\sim$26.4 ksec) missed 9\% of the orbit just prior to eclipse
ingress. The counting statistics of the RGS instruments were too low,
and we dismissed those spectra from further analysis.  Data were
processed with version 5.3.1 of the XMM-{\ii Newton} Science Analysis
Software (SAS). Source photons were extracted with{\ii in}
aperture{\ii s} of 45\arcsec\ and 30\arcsec\ {\ii radius} around the
position of \ob\ for the pn and MOS data, respectively. The background
which was {\ii at} a low and constant level during the pointing was
determined from a source free area close to the target.  For spectral
analysis within XSPEC, only single and double events were used
together with the appropriate response files pn\_ff20\_sdY9 and
m[1,2]1\_r7\_im\_all\_2000-11-09.  For the pn detector, operated in
full window mode, the fitting procedure was complicated by photon
pile-up for count rates {\ii exceeding} 5 cts~s$^{-1}$. We tried to
minimise these effects by selecting only photons from either low count
rate intervals or an outer annulus of the PSF.

In order to improve the counting statistics and {\ii to provide a}
compact display, we constructed combined light curves from all EPIC
instruments with the count rates of the two MOS CCDs normalised to the
level of the pn detector.  EPIC photon timings given in the time frame
of terrestrial time (TT) have been converted to the barycenter of the
Sun.  We have adopted the orbital ephemeris of \cite{Staude01}, which
uses the mid-eclipse of the white dwarf as \mbox{zero-point}.  The
corresponding value $244\, 50339.43503$ in the time frame of
barycentric Julian ephemeris days (BJED) has been corrected for a
typing error and the appropriate number of leap seconds.

Simultaneously to the X-ray observations, \ob\ was observed with the
optical monitor (OM) in the UVW2 (1800-2400 \AA) band.  The OM was
operated in image mode resulting in $23\times 800$~sec
integrations. We applied aperture photometry to the images provided by
{\ii the} {\verb omichain } task to derive background subtracted count
rates of the source. These were converted to flux units using {\ii
the} conversion factor $5.242\times 10^{-15}$ \erga\ {\ii cts$^{-1}$}
given in the XMM{\ii -Newton} documentation.

We also included {\ii in our analysis} 34 ksec of largely unpublished
ROSAT data taken between 1991 and 1996 with the PSPC and HRI
instruments. These observations {\ii we}re spread into small blocks
separated by days or weeks, which cover only small fractions of the
orbit at a given epoch.  Photon event files have been taken from the
ROSAT archive and standard corrections (vignetting, dead-time) {\ii
have been applied} using the EXSAS software package
\citep{Zimmermann94}. 
{\ii For comparison with the PSPC data, HRI count rates have been multiplied 
with a factor of 6.}

\section{X-ray and UV light curve}
\subsection{Flare phase}
\label{s:flarephase}
In Fig.~\ref{f:lcall}a we show the XMM{\ii -Newton} EPIC light curve
folded over the orbital ephemeris and binned at a resolution of 30
sec.  The most prominent feature is a short, bright interval {\ii at}
\phiorb = 0.41--0.68 with peak count rates reaching 15 cts/sec (or
25 cts/sec at a binning of 3 sec).  At higher temporal resolution
(Fig.~\ref{f:flare}) most of the bright phase flux can be resolved
into individual flares.  The {\ii frequency} and intensity of the
flares strongly varies with the largest flares seen shortly after
onset of the bright phase and a smoothly declining intensity
thereafter.

\begin{figure}[t]
\includegraphics[clip=,angle=-90,width=0.99\columnwidth]{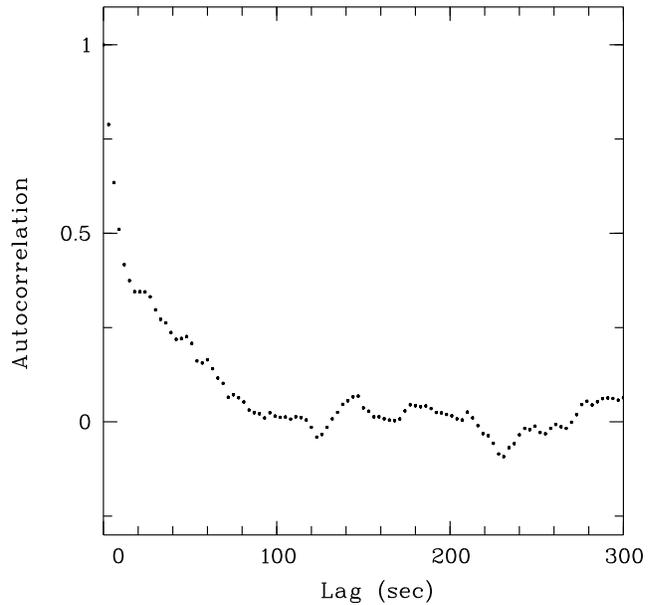}
\caption{ Auto-correlation {\ii function} of the bright-phase light
curve.  Time delay bins are 3 sec.  }\label{f:auto}
\end{figure}

Most of the individual flares are well-separated with average rise and
decay times of $\sim$10~sec.  This time-scale {\ii corresponds to} the
first part of the auto-correlation function {\ii which} steeply falls
off with an e-folding time of 7\,s {\ii (Fig.~\ref{f:auto})}.  The
correlations remain positive for larger lags up to {\ii the}
zero-crossing time {\ii of $\sim 100$} s, which is interpreted as the
average time between two consecutive flares.  For a dozen well
isolated flares we measured the integrated fluxes as well as the
corresponding peak flux. This sample includes a characteristic set of
faint and bright events ranging between $F_{\rm int} =
(0.6-3.2)\times10^{-9}$erg~cm$^{-2}$ and $F_{\rm peak} =
(2-7)\times10^{-10}$\ergcm and should be representative for the
majority of flares seen {\ii in} Fig~.\ref{f:flare}.

Perhaps the greatest surprise {\ii are} the spectral propert{\ii ies}
of the flaring emission best seen in the hard X-ray light curve taken
in the {\ii 1--8 keV band} (Fig.~\ref{f:lcall}b). While a constant
source of faint, hard X-ray flux (see Sect.~\ref{s:faint_lc}) {\ii is}
observed at all phases apart from the eclipse, there are no X-ray
photons with energies larger than 1 keV that can be uniquely
attributed to the soft X-ray flares.  {\ii We derive a}n upper limit
{\ii for} the hard X-ray flux from the flares {\ii using} the counting
statistics of the residual faint flux. In the bright{\ii -phase}
interval $\phiorb =$ 0.43--0.6, 650 source and background photons
with energies $> 1$~keV {\ii have been detected. This corresponds} to
a detection limit {\ii of} 25 photons or 0.0033 cts/s {\ii for the
hard X-ray count rate related to the soft X-ray flares}.  A similar
conclusion is {\ii reached} if {\ii we assume that} the 25 photons
from the hard mini-flare observed at \phiorb = 0.46 {\ii are
associated with} a soft flare event.  {\ii Assuming a} bremsstrahlung
spectrum {\ii with $kT_{\rm br}$ =} 20 keV {\ii the upper limit to
the} count rate {\ii is} translate{\ii d in}to a bolometric flux limit
$F_{\rm br, bol} \leq 2.7\times 10^{-14}$ \ergcm .  We also supply the
equivalent value for the ROSAT band $F_{\rm br, 0.1-2.4 keV} = 5.6
\times 10^{-15}$ \ergcm\ to aid comparison with the studies of
\cite{Beuermann95}.

A few isolated soft flare events are observed outside the bright{\ii
-phase} interval. For example, there is a series of six mini-flares
seen shortly prior to the eclipse at \phiorb = 0.94 for a duration of
400 sec.  The highly instationary nature of the accretion process
makes it difficult to clearly differentiate between geometrical
aspects and instantaneous variations of the mass transfer rate.  The
smooth UV light curve indicates that \ob\ was permanently accreting at
a high rate and we conclude that the entire bright interval was not
due to an isolated, long-lasting accretion event.  {\ii Assuming that}
the duration $\Delta\phi = 0.27$ and the center of the bright phase
$\phi = 0.545$ {\ii reflect the geometrical constraints on} the
visibility of the accretion region, would imply an {\ii unusual
location of this region at} an azimuth $\psi\sim 160\degr$ and a
colatitude $\beta\sim 160\degr$ (see discussion in
Sect.~\ref{s:geom}).

Additional information {\ii on} the {\ii duration and the center of
the phase interval during which strong flares occur} is provided by
various ROSAT light curves (Fig.\ref{f:lcall}c) of \ob\ obtained from
pointed observations at different epochs.  Compared to the XMM{\ii
-Newton} pointing, flare intervals are distributed over a larger phase
range $\phiorb =$ 0.45--0.87 and shifted towards phase zero.  The
brightest episodes are observed around $\phiorb = 0.8$, i.e.  in the
faint{\ii -phase} interval of the XMM{\ii -Newton} observation
indicating a possible migration of the accretion spot.  The general
occurrence of the bright intervals is consistent with an accretion
region located at an azimuth $\psi\sim 90\degr$ and {\ii at a}
colatitude $\beta > 90\degr$.  {\ii All available ROSAT and XMM-Newton
data agree that flares do not occur in the} post-eclipse interval,
$\phiorb =$ 0.05--0.45.

\begin{figure}[t]
\begin{center}
\includegraphics[clip=,width=0.90\columnwidth,angle=-90]{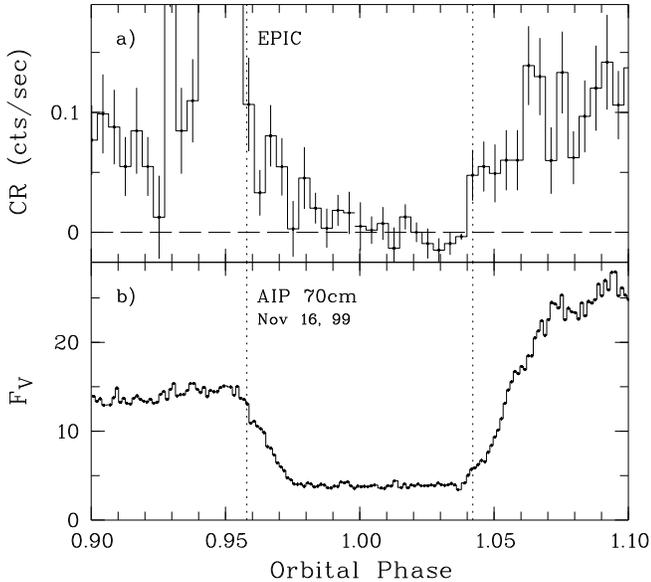}
\caption{ {\ii a) EPIC pn/ MOS X-ray eclipse light curve} of \ob\ {\ii
binned} at {\ii a time} resolution of 240 s {\ii and b)} optical light
curve {\ii obtained with the 70-cm reflector of the AIP shown for}
comparison. The vertical lines mark the {\ii orbital phases of the}
white dwarf ingress/ egress measured from UV HST light curves
\citep{Staude01}.  }
\label{f:lcecl}
\end{center}
\end{figure}

\begin{figure}[t]
\begin{center}
\includegraphics[clip=,width=0.98\columnwidth]{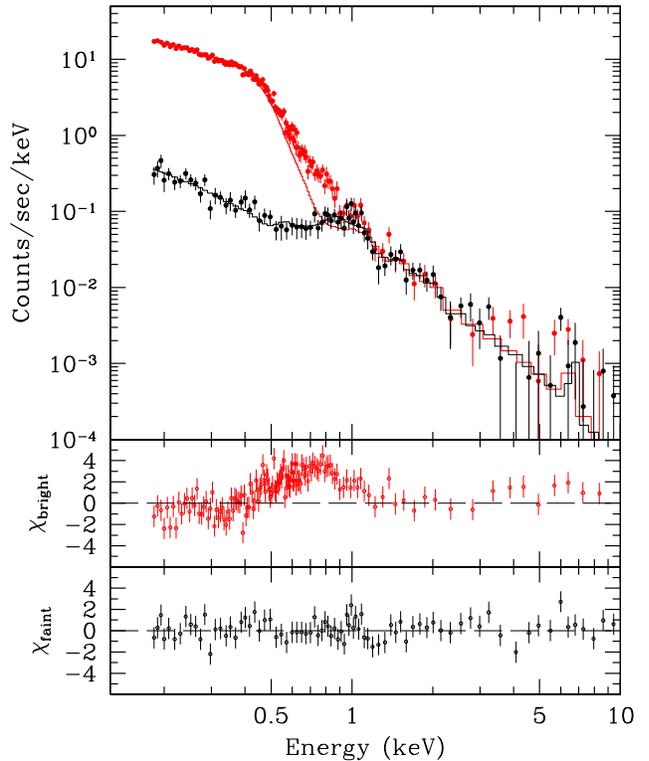}
\caption{ EPIC pn spectra of \ob\ taken during bright and
faint phase, together with likely model spectra and their
residua.  The model of {\ii the} faint phase spectrum is a
3-temperature {\it mekal} fit.  The bright phase spectrum shown here
includes piled-up photons, while the fit to the data uses fixed
parameters derived from the pile-up free spectrum.}
\label{f:xspec}
\end{center}
\end{figure}

\subsection{Faint phase and eclipse profile}
\label{s:faint_lc}
Throughout the orbit we detect a persistent, faint ($\sim$0.1cts~s$^{-1}$)
source of emission 
(Fig.~\ref{f:lcall}b).  {\ii The spectrum of} this emission is
significantly harder {\ii than the spectrum of the flaring component
and} about 50\% of all photons {\ii are} detected at energies greater
than 1 keV.


During the eclipse by the secondary star 
the source of the faint emission is occulted {\ii
(Fig.~\ref{f:lcecl})}.  The beginning of X-ray ingress coincides with
that of the white dwarf observed at $\phiorb = 0.976$ in the UV
\citep{Staude01,Schmidt01}.  Surprisingly, the ingress is prolonged to
phases {\ii beyond the geometrical occultation of} the white dwarf.
{\ii In total, a significant} excess emission of $28\pm 7$ photons
{\ii has been detected during the eclipse} above {\ii the} background
{\ii level} of one photon.
A rough estimate of the spectral characteristics of the ingress flux can 
be made from the ratio of photons above and below 1 keV which is  0.5. 
From spectral analysis (Sect.  \ref{s:faint_spec}) we find a 
similar value 0.36  for {\ii the}
intermediate temperature (0.7 keV) component of the faint emission.

The estimated egress duration of 15--20 minutes is much longer than
$\sim$45~sec expected for the occultation of an 0.7\Msun\ white dwarf,
and is difficult to reconcile with a compact emission region.  A
similar, corresponding behaviour for eclipse egress is hard to
confirm, due to the low counting statistics and the intrinsic
variability of the source.


\subsection{The UV light curve}
The UV light curve shown in Fig.~\ref{f:lcall}d is similar to the
optical light curves \citep{Shafter95} with a primary maximum around
$\phi\sim 0.2$ and a secondary brightening at phase $\phi\sim
0.6$. Such asymmetric double-humped shape is interpreted as the result
of the changing projection of the accretion flow.  During the rise of
the X-ray bright phase the UV flux is actually decreasing, indicating
that the accretion spot itself contributes only a minor fraction in
the UV range.  With the sparse sampling, eclipse ingress and egress
are not resolved. For the single OM exposure centred at $\phi\sim$ 0,
the eclipse is total.  The flux levels before and after eclipse are
comparable to that of a{\ii n} HST FOS observation of \ob\ taken in
1996 \citep{Staude01}.  We conclude{\ii , therefore,} that \ob\ was in
a normal high state {\ii of accretion} during the XMM-Newton
observation.

\section{X-ray spectroscopy}
\subsection{Flare phase}
\label{s:bright_spec}

   \begin{table*}
     \caption{Fit results for faint and bright phase X-ray spectra of \ob}
      \begin{tabular}{lllccccccl}
      \hline
      \noalign{\smallskip}
      Data & Model & $N_{\rm H}$ 
                  & $kT_{\rm bb}$ 
                  & $kT_{1}$ 
                  & $kT_{2}$ 
                  & $kT_{3}$ 
                & $\alpha$
                & $F_{\rm bol}$
                & $\chi^2_{red}$ \\
         & &  (10$^{20}$ cm$^{-2}$)  & 
         (eV) &
         (eV) &
         (keV) &
         (keV) & &
         $({\rm erg~cm}^{-2}{\rm s}^{-1})$  & \\
       \noalign{\smallskip}
      \hline
      \noalign{\smallskip}
{\it Faint phase }  & & & & & & &\\[3pt]
pn & bb+br    & $5.1$   & 32$\pm3$&--&1.2$\pm0.3$&--&--&$1.2\times10^{-12}$& 1.4\\
pn & bb+mek   & $<0.01$ & 68$\pm3$&--&1.1$\pm0.1$&--&--&$0.3\times10^{-12}$& 1.8\\
pn& 2mek&$<0.01$&--&$80\pm5$&2.7$\pm0.5$&--&--&$2.5\times10^{-12}$& 1.7\\
pn& 3mek& $2.5$   &--&$65\pm2$ & 0.7$\pm0.2$ & 2.9$\pm 0.4$ && 
         $1.5\times 10^{-11}$ & 0.96\\
pn & cmekal & $<0.01$  &--&--&--& 7.1$\pm 0.4$&0.7$\pm0.2$ &
         $4.3\times10^{-13}$ & 3.1\\
pn & bb+cmekal &$1.4$  &47$\pm7$&--&--& 5.5$\pm 0.7$ &1.0$\pm0.2$& 
         $7.2\times10^{-13}$  & 0.86\\

 & & & & & &  & & &\\
{\it Bright phase }  & & &  & & & & &\\[3pt]
MOS1    & bb+3mek
           & 8.1$\pm$0.3  & 42$\pm4$ &65\ste&0.7\ste     & 2.9\ste  &--&
  $18.9\times 10^{-11}$ & 0.87\\
MOS2    & bb+3mek
           & 6.0$\pm$0.3  & 42$\pm3$ &65\ste&0.7\ste     & 2.9\ste  &--&
  $16.8\times 10^{-11}$ & 0.97\\
pn (outer psf) & bb+3mek
           & 5.0$\pm$0.2  & 46$\pm2$ &65\ste&0.7\ste     & 2.9\ste  &--&
  $7.48\times 10^{-11}$ & 0.86\\
pn (CR$<2$ s$^{-1}$) & bb+3mek
           & 3.7$\pm$0.7  & 44$\pm2$ &65\ste&0.7\ste     & 2.9\ste  &--&
  $7.43\times 10^{-11}$ & 1.04\\

\noalign{\smallskip} \hline \noalign{\smallskip}
\end{tabular}

\noindent{\small 
\ste\ These parameters were fixed. 
}
   \label{fitres}
   \end{table*}

The spectrum of the {\ii X-ray emission seen in the} flar{\ii ing}
phase can be {\ii adequately} fitted by a single blackbody and an
extra component representing the unrelated hard {\ii X-ray} emission
(Fig.~\ref{f:xspec}). For the latter we have adopted the 3-temperature
{\tt mekal} model determined in Sect.~\ref{s:faint_spec} with fixed
best-fit parameters.  The blackbody temperatures derived from the {\ii
data of the} two MOS CCDs, {\ii $kT_{\rm bb} = (42\pm 4)$}~eV, and
from the pile-up free pn spectrum, {\ii $kT_{\rm bb} = (45\pm 3)$}~eV,
agree within the error bars {\ii (Table~\ref{fitres})}.  The {\ii
temperature is comparable with values observed in other polars with a
pronounced blackbody component, which roughly cover the range of
20--60 eV}. The hydrogen column density of
$(5-8)\times$10$^{20}$\,cm$^{-2}$ towards \ob\ is in agreement with previous 
ROSAT and BeppoSAX measurements \citep{Beuermann95,ElKholy04,deMartino98a} 
and slightly lower than the total interstellar galactic column of
10$^{21}$cm$^{-2}$ in this direction.  The {\ii remaining
uncertainties in} $T_{\rm bb}$ and $N_{\rm H}$
{\ii imply a corresponding uncertainty in the unabsorbed bolometric
soft X-ray flux $F_{\rm bb,bol}$, which averages} $17.5\times
10^{-11}$ \ergcm\ for the two MOS CCDs and $7.8\times 10^{-11}$
\ergcm\ for the pn.  In addition to calibration issues of the low
energy detector response, the pn flux may still be somewhat influenced
by the photon pile-up and we will adopt the MOS flux for later
considerations.

We have searched for possible change{\ii s} of the blackbody
temperature as a function of count rate.  Bright phase X-ray spectra
selected for intervals with count rates lower or higher than 5
cts~sec$^{-1}$ have indeed different parameters, but these variations
are only significant at the 3$\sigma$ level and not consistent between
the different detectors. For the MOS1 and {\ii the} pn CCD, the low
count rate spectra have lower temperatures of 35 {\ii and 42 eV,
respectively,} compared {\ii with} $\sim$45~eV for the bright flare
measurements with all instruments. The fits of the MOS2 {\ii data} on
the other hand differ only {\ii in} the hydrogen column {\ii density}.

\subsection{Faint phase}
\label{s:faint_spec}
The faint phase spectrum was extracted for phase ranges $\phi_{\rm
orb} =$ 0.03--0.41 and $\phi_{\rm orb} =$ 0.68--0.97 carefully
excluding {\ii the} eclipse interval as well as a few intermittent
flares. Since the signal-to-noise {\ii ratio} of the MOS faint phase
spectra was too low to constrain different models only the pn data
were considered further.  The spectrum shown in Fig.~\ref{f:xspec} is
much flatter and harder compared {\ii with that of} the bright phase,
and can {\ii not} be fitted by {\ii (i)} a single blackbody{\ii ,
(ii)} a {\ii single temperature optically thin thermal emitter, and
(iii) a combination of both.}
The {\ii spectral shape rather suggests an optically thin plasma
emitter with a range of comparatively low temperatures. A good fit was
obtained with Mewe models
({\verb mekal }, \citealt{mewe85}) with temperatures $kT_{\rm mek} = 0.065$,
0.7, and  2.9 keV{\ii, yielding }} 
$\chi^{2}_{\rm red} = 0.95$.  The total bolometric flux {\ii of the
faint emission is} $F_{\rm faint} = 1.42\times 10^{-11}$ \ergcm ,
with the largest fraction ($\sim$ 90\%) coming from the component with
the lowest temperature.  Since the emission measure in the cooling
flow of a bremsstrahlung-dominated plasma shock-heated to $T_{\rm
max}$ varies approximately as $T/T_{\rm max}$ it is obvious that the
parameter set for a single cooling flow can not match the
observation. Attempts to fit the faint-phase emission with the
multi-temperature model ({\verb cmekal }, \citealt{Done98}), in which
the emission measure for a given temperature is scaled by the
expression $(T/T_{\rm max})^{\alpha}$, were in fact unsuccessful
($\chi^{2}_{\rm red} \ge 3.1$ for any possible combination of $T_{\rm
max}$ and ${\alpha})$. 

{\iii 
The residuals for {\iiii this} model are confined to 
energies below 
0.5~keV, indicating the presence of an independent low temperature component.
Adding a blackbody component to the} {\verb cmekal } {\iii model yields an
acceptable fit $(\chi^{2}_{\rm red} = 0.86)$, but also an optically thin 
thermal model cannot be excluded. The temperature of the blackbody 
is {\iiii $(47\pm7)$} eV and agrees within the errors with that of the primary soft
bright phase emission. 
}


\section{Discussion}

\subsection{The accretion rate of \ob}

The bolometric time-averaged bright-phase luminosity of the flare
emission is $L_{\rm bb} = C\pi F_{\rm bb} d^{\,2} = 2\,C\times
10^{33}$\ergsec\ calculated for a distance of 625~pc\footnote{We
update the distance estimate by \cite{Staude01} using the more
accurate Hipparcos distance of the comparison star YY Gem.}.  With
$C=1$, this luminosity refers to the emission received from a plane
surface element viewed face-on. If the plane surface
element is located at co-latitude $\beta$ in a system seen at
inclination $i$, \mbox{C=1/cos$(\beta -i)$} at the best visibility of
the element and larger if closer to the limb of the star. With $\beta
= 160\degr$ and $i = 78\degr$ \citep{Staude01}, a correction by a
factor as large as $\sim7$ would be required. There is evidence,
however, from various studies that the blackbody emission arises from
thermally or dynamically elevated mounds of hot photospheric matter
rising up to a couple of percent of the white dwarf radius
\citep{Schwope01}. In their early study, \citet{Heise85} noted that
the individual (and transient) mounds are about as high as wide. As a
compromise, we use a factor $C\simeq 2$, which corresponds to
isotropic emission into the half sphere. The soft X-ray blackbody
luminosity then is $L_{\rm bb}\simeq 2\pi F_{\rm bb}
d^{\,2}$. Neglecting, for simplicity, any additional contribution to
the accretion luminosity, e.g., from cyclotron radiation, $L_{\rm bb}$
corresponds to a time-averaged accretion rate $\dot M = L_{\rm
bb}R/{\rm G}M$. Assuming a standard white dwarf of mass $M=0.7$ \Msun\ 
and radius $R=8\times 10^8$\,cm, we obtain $\dot{M} =
2.7\,C\times10^{-10}$\,\Msun yr$^{-1}$. For $C=2$, $\dot{M} =
5.4\times10^{-10}$\,\Msun yr$^{-1}$, comparable to the $\sim 9\times
10^{-10}$\,\Msun yr$^{-1}$ estimated by \citet{Beuermann95} from the
ROSAT All-Sky-Survey observation of \ob\ if one corrects for their
lower (assumed) value of $T_{\rm bb}$. It is also close to the
$\sim 6\times 10^{-10}$\,\Msun yr$^{-1}$ derived by \citet{ElKholy04}
from the ROSAT pointed observation. Our value of $\dot M$, however, falls
substantially below the rates which seem to be typical of dwarf novae
and novalike variables as estimated by \citet{Patterson84} and implied
by the effective temperatures of accreting white dwarfs if interpreted
in terms of compressional heating \citep{Araujo05,Townsley04}. This
seems to support the notion that long-period polars have lower
accretion rates than other CVs \citep{Araujo05}.

\subsection{The case for pure blobby accretion}

Our XMM -Newton observation revealed \ob\ as extreme among polars
in its dominance of soft over hard X-ray emission. From the bolometric
flux $F_{\rm bb}$ of the soft X-ray blackbody and the upper limit to
the bremsstrahlung flux $F_{\rm br}$ (Sect. \ref{s:flarephase}) we
derive a lower limit to the soft-to-hard bolometric X-ray flux ratio,
$F_{\rm bb}/F_{\rm br} \geq 6700$.  An even larger value, $F_{\rm
bb}/F_{\rm br} \geq 25000$, is obtained if the fluxes are restricted
to the ROSAT band \citep{Beuermann95}.  No other polar
shows such a weak hard X-ray component.
The best explanation for the lack of hard X-ray emission in
polars is the accretion of blobs, or rather field-aligned filaments, of
matter which penetrate to sub-photospheric layers releasing their
energy in shocks which are entirely submerged in the photosphere
\citep{Kuijpers82,Frank88,Beuermann04}. Shock heating of the infalling
matter can not be avoided in the super-sonic flow, but burying of the
shock and complete thermalization can render the hard X-rays
undetectable.

As noted already by \citet{Walter95}, the soft X-ray emission in \ob\
consists of individually identifyable flares. The spectral properties
of \ob\ suggest a high density of the infalling matter and the
temporal structure of the emission suggests the infall of blobs or
filaments of matter. The detection of individual flares provides the
opportunity to estimate the physical parameters of these
filaments. While several causes may be responsible for the high
density of the matter, an obvious one is the compression of the
filaments perpendicular to the field as $r^{-2.5}$, with $r$ the
radial distance \citep{Beuermann87}. The surface field strength $B$ in
the accretion spot has been measured by \citet{Shafter95} and
\citet{Schmidt04} from the spacing of the, admittedly weak, cyclotron
harmonics, yielding $B=61$\,MG. 

The stagnation radius obtained by equating the pressure of the equatorial
magnetic field and the free-fall ram pressure of the accretion stream 
expressed in units of the white dwarf radius is
\begin{eqnarray*}
     r_{\rm s}/R \simeq 7.2 B_7^{2/5} n_{14}^{-1/5} R_9^{1/5}
     \left(M/M_{\odot}\right)^{-1/5}
\end{eqnarray*}
\citep{Beuermann87}, where $B_7$, $n_{14}$, and $R_9$ are the polar field
strength, the electron density near the stagnation point, and the radius
of the white dwarf in units of $10^7$\,G, $10^{14}$\,cm$^{-3}$, and 
$10^9$\,cm, respectively.  Starting from a typical electron density in 
the threading region $n_{14} =$ 0.1--10 we obtain $r_{\rm s}/R =$ 10--24 
for $B=61$\,MG and a white dwarf of 0.7 \Msun . The pre-shock density at 
the white dwarf is a factor of $\left(r_{\rm s}/R\right)^{2.5} =$ 300--2800 
times higher and the mass density becomes $\rho_0\simeq (0.6-6.0)\times 
10^{-7}$\,g\,cm$^{-3}$.  The specific accretion rate then is $\dot m \simeq$ 
28--280\,g\,cm$^{-2}$s$^{-1}$. This is much higher than the value $\dot m 
\simeq 3$\,g\,cm$^{-2}$s$^{-1}$ \citep[ his Fig.~4]{Beuermann04}, at which 
the shock is depressed to the level of the photosphere. Hence, we expect 
the shocks associated with the infall of the flare-producing blobs in \ob\ 
to be deeply buried in the atmosphere of the white dwarf.

{\iii
In addition, 
the production
of hard X-rays from the low density part of the $\dot{m}$-distribution 
is {\iiii diminished by} the dominance 
of cyclotron emission over thermal bremsstrahlung as the primary coolant 
in a high field plasma \citep{Woelk96}{\iiii , like in \ob{}}. 
Furthermore, the viewing geometry towards moderately buried shocks may 
play a role.
Reprocessing of hard X-rays occurs already for moderate suppression of
the shock if the accretion spot is located sufficiently close to the
limb of the star. For a geometry as in \ob\ with a minimum viewing
angle $\beta - i \sim 80\degr$ the actual path length through
atmospheric matter will exceed that for a face-on view by a large
factor and shield shocks that would be directly visible in a face-on
view \citep{Frank88,Beuermann04}.  
Another constraining parameter is the number of blobs simultaneously
impinging on the white dwarf. The visibility of shocks will also be
influenced by the interaction of the individual impact spots and the
splashes caused by them, an effect which increases with the number N
of such impact regions present simultaneously. The structured X-ray
light curve of \ob\ suggests that this number is small, of the order
of unity, and certainly less than estimated for AM Her by
\citet{Hameury88}.
}

Several complications may disturb the simple picture presented above.
E.g., the density in the threading region at the start of the
quasi-free fall may vary between different objects because it depends
on the balance between cooling and heating experienced by the matter
in the accretion flow, processes which are difficult to assess
observationally and theoretically. Break-up of the stream into
separate blobs may occur already near the inner Lagrangian point or be
effected by instabilities in the magnetosphere \citep{Hameury86}. As a
tendency, we presume that the extended path lengths in the long-period
system \ob\ favour cooling and higher densities.


While the ratio $F_{\rm bb}/F_{\rm br}$ tends to increase with the
magnetic field strength $B$ \citep{Beuermann94,Ramsay94}, there is no
simple one-to-one relationship. In the polar sample studied during
the ROSAT All-Sky-Survey, the highest ratios $F_{\rm bb}/F_{\rm br}
\sim 100$ are observed for systems with a field strength similar to
that of V1309 Ori \citep{Beuermann95}. There are some exceptions,
however, suggesting that compression in the magnetic field is not
sufficient to suppress all hard X-ray emission. The complex
observational relation between $F_{\rm bb}/F_{\rm br}$ and $B$ is most
easily explained if there are sources of keV X-rays which are not
associated with buried shocks. Such source is present also in \ob\ and
is addressed in the Section 5.4.


In summary, we consider it plausible that a number of reasons combine
to cause the observed dominance of soft X-rays observed in \ob.

\subsection{Blob properties}

In contrast to many other polars, the number of simultaneously
impacting filaments is small and a large fraction of the soft X-ray
flares can be individually resolved, thus providing imminent clues on
the temporal behaviour and the impact energetics.  On average, the
flare profiles are symmetric with rise and decay times of about
10~sec. The lack of obvious temperature changes throughout individual
bursts suggests that the intensity profile of the flare is determined
by a corresponding profile of the local mass flow rate per unit area
and instantaneous thermalization. A finite cooling time of the heated
surface elements is not discernible in the data. The individual flares
in \ob\ release energies in the range of $(1-6)\times10^{34}$ erg,
which we estimate using a distance of 625~pc and a geometry factor of
$2\pi$.  For a white dwarf of 0.7\Msun , this range of energies
corresponds to a range of blob masses $m_{\rm
blob}=(1-6)\times10^{18}$g.

At flare maximum, the blackbody emitting areas $A_{\rm bb}$ computed
from
\begin{eqnarray*}
     L_{\rm eff,peak} = A_{\rm bb,peak} \sigma T^{4}
\end{eqnarray*}
are typically in the range of $(0.6-1.8)\times10^{15}$cm$^{2}$.  With a
typical flare duration $\Delta t\sim 10$~sec, the characteristic
length of an accreted filament is $l\simeq v_{\rm ff} \Delta t\simeq
4\times 10^{9}$~cm. Combined with the emitting area $ A_{\rm
bb,peak}$, we deduce an upper limit for the volume of the infalling
filament of \mbox{$V< A_{\rm bb,peak}\times l$}, where the limit derives from
the fact that the area over which accretion occurs $ A_{\rm
acc}<A_{\rm bb,peak}$ since the heat flow will spread sideways from
the column. For a typical flare produced by a blob mass $m_{\rm
blob}=2\times 10^{18}$g, the lower limit on the density of the infalling
matter before shock compression is
\begin{eqnarray*}
 \rho_0 > m_{\rm blob} l^{-1} A_{\rm eff,peak}^{-1} \simeq 8\times
 10^{-7}{\rm g\,cm^{-3}}
\end{eqnarray*}
which translates into a specific accretion rate 
\begin{eqnarray*}
 \dot m=\rho_0 v_{\rm ff} > m_{\rm blob} v_{\rm ff}\,l^{-1} A_{\rm
 eff,peak}^{-1} \simeq 300 {\rm g\,cm^{-2}s^{-1}},
\end{eqnarray*}
about two orders of magnitude higher than the $\dot m$ at which the
shock is depressed to the level of the photosphere (Beuermann
2004). Hence, the hard X-ray emission region is likely to be
completely engulfed by atmospheric and infalling matter and hard
X-rays are suppressed at practically any viewing angle. This limit on
$\dot m$ of the flare-producing accretion events compares favourably
with the value estimated above from the density in the threading
region and the compression in the magnetic funnel.

\subsection{Origin of the faint-phase emission}

Besides the blackbody-like soft X-ray component visible only 
during the flares, significant excess emission is detected at energies 
up to 3 keV during almost all orbital phases.  The spectral and temporal 
behaviour of this faint-phase emission is complex and atypical of the 
hard X-ray emission which is seen from other magnetic CVs and usually 
attributed to bremsstrahlung from the accretion shock.  The shock 
temperature is expected to be \citep{Aizu73}
\begin{eqnarray*}
kT_{\rm shock} = \frac{3}{8} \frac{GM}{R} \mu {\rm m_{\rm u}} = 
32 \frac{M}{M_{\odot}} \frac{1}{R_9} {\rm keV},
\end{eqnarray*}
where $m_{\rm u}$ is the unit mass and $\mu= 0.617$ for solar composition 
has been used.  The absence of a bremsstrahlung component with 
$kT_{\rm shock} \sim$\,20 keV is compliant with our finding that in \ob\ 
the shocks in the accretion plasma are deeply buried in the white dwarf 
photosphere and a different origin is required for the two components 
distinguishable in the faint-phase emission.

{\it 1) Soft component:}  
The softest part with plasma temperature $kT_{\rm mek} \sim 65$ eV 
comprises the largest fraction of the bolometric flux of the faint-phase 
emission.  In contrast to the hard X-ray flux, this component shows an 
orbital modulation with increasing flux towards the bright phase.  The 
similar temperature and orbital variations of the soft component and the 
flaring component suggest that both have a similar origin from the primary
accretion region.  Scattering and reflection of soft X-rays in the 
accretion column above the impact site is a viable mechanism that could 
explain the basic spectral and temporal properties of the soft component
in the faint phase.  For any accretion geometry it can not be avoided 
that part of the photons from the blackbody-like soft X-ray component 
are scattered in the infalling material.  As this material extends to 
several white dwarf radii, scattered X-ray photons will still be visible 
when the accretion region itself is occulted by the white dwarf.  
Scattered emission is largely outshone in systems which show a strong 
bremsstrahlung component and it is, therefore, not surprising to find 
such a component specifically in \ob .

{\it 2) Hard component:}
About 10\% of the bolometric flux of the faint-phase emission arises
from an additional hard X-ray component with $kT_{\rm mek} \sim $1--3 keV
showing no significant orbital variation besides a gradual decrease during
the eclipse by the secondary star.  Ingress appears to be prolonged
after the nominal eclipse of the white dwarf.   The estimated duration of 
the ingress is 15--20 minutes and corresponds to a linear extent of 
$3.2\times10^{10}$~cm or 40\,$R$ at the location of the white dwarf.  The 
comparably low plasma temperature and the large geometrical extension of
this emission component provide strong arguments against an origin from a 
compact accretion region.

A possible source for the observed emission could be plasma heated by
shocks in the coupling region where the ballistic stream is
captured by the magnetic field. At the typical coupling radius of 10--30
$R$ the kinetic energy of the free-falling material amounts to 3-10\,\%
of the total gravitational energy and a certain, but yet unknown fraction 
of this energy is likely to be dissipated at that point.  If a shock 
arises at $r_{\rm s} =$ 10--30 $R$, a shock temperature in the range 
0.9--2.8 keV is expected in agreement with our X-ray spectroscopy.  The 
bolometric luminosity of the hard component is $1.4\times 10^{31}$~\ergsec\, 
for a distance of 625 pc and a geometry factor of $4\pi$.
For a mass accretion rate $\dot{M} = 5.4\times10^{-10}$~\Msun~yr$^{-1}$, 
an accretion luminosity in the range of (1.3--4.0)\,$\times 10^{32}$~\ergsec\  
is expected at a radial distance $r_{\rm s} =$ 10--30~$R$, demonstrating 
that there is no energy problem in providing the observed faint-phase X-rays.

\subsection{Accretion geometry}
\label{s:geom}

Modelling of low level variations of the optical circular polarisation 
indicates a two-pole geometry \citep{Buckley95,Katajainen03} with the two 
poles located at $(\phi_{1}, \beta_{1}) = (-70\degr, 145\degr)$ and 
$(\phi_{2}, \beta_{2}) = (110\degr, 35\degr)$.  Because of the strong
stream emission, however, the polarised emission is faint and conclusions
on the accretion geometry should be considered with some caution.

Our X-ray observations provide some additional insight.  The flaring 
soft X-ray emission seen between phases 0.4 and 0.95 in the ROSAT and {\iii 
XMM data are in general agreement with the azimuth of the 
second optical spot. }
{\iii The large variation {\iiii of the phase intervals during which}
soft X-ray emission {\iiii is seen at different epochs}
indicates either real changes of the accretion geometry or
fluctuations in $\mdot$.  
Th{\iiii e latter} possibility is supported by
small scale flaring which continues between phases 0.7 and 0.95 
during the XMM-Newton pointing.  
{\iiii If} 
on the other hand, the XMM-Newton bright phase between $\Phi =$ 0.41--0.68
{\iiii indicates}
the location of the X-ray emitting region 
then its position would be quite peculiar. 
The centre of th{\iiii is} bright interval {\iiii would be}
consistent with {\iiii a spot at} an azimuth $\psi = 160\degr$ {\iiii which is} 
almost antipodal to the {\iiii direction towards the} secondary star
($\psi = 0\degr$).  
Only the polar HY Eri \citep{Burwitz95} has a main accretion region 
at such extreme position, whereas the primary accretion pole seems to 
be generally confined in the range $0 - 60\degr$ \citep{Cropper88}. 

}

Because of the pronounced time variability a
final conclusion cannot be drawn.  The second optically derived accretion
spot, $(\phi_{1}, \beta_{1})$, is visible after the eclipse during orbital 
phases 0.05--0.39.  In principle, this pole could be responsible for the
post-eclipse faint-phase X-ray emission.  The spectral properties of
this component do not correspond to the known characteristics of shock 
emission, however.  We consider, therefore, the scattering scenario 
described above as more likely and conclude that the second optical pole
is not seen in X-rays.

\section{Conclusions}

{\iii
 Our XMM-Newton and ROSAT observations of \ob\ have revealed 
 the most extreme example of a soft X-ray excess yet found in any magnetic 
 CV. Our density estimate and constraints from the observation itself
 show that the bulk of the mass flow will be buried 
 to sufficiently deep sub-photospheric layers. Other 
 mechanisms (cyclotron cooling, geometry) may be responsible 
 for the quenching of additional hard X-ray emission 
 from any residual low density material. 

The soft X-ray emission is seen 
between orbital phases 0.4 and 0.95 consistent with an origin from one
of the polarimetrically defined accretion spots.  
Further observations that will provide a much better definition of the 
mean orbital light curve are required to clarify the yet uncertain 
accretion geometry of \ob . 


%

{\iiii
The softness of the X-ray emission of the primary accretion region and
its favourable geometry 
allowed us to disentangle two other emission components which have not been 
resolved in any other polar so far. The softer of these components has been  
interpreted as scattering or reflection of primary soft X-ray photons in 
the column closely above the impact region. 
}

The prolonged eclipse on the other hand, indicates {\iiii the presence of}
a much more extended emitting source for the harder faint phase emission,
most likely the accretion stream.  {\iiii A possible mechanism for this 
component is shock heating at the stagnation region in the magnetosphere.  It 
appears, however, that further progress and a definite conclusion on this
interesting possibility require a much better definition of the 
mean orbital light curve and an improved statistical definition of the 
decreasing X-ray emission during the eclipse.
}

}

\begin{acknowledgements}
RS was and is supported by the Deut\-sches Zentrum f\"ur
Luft- und Raumfahrt (DLR) GmbH under contracts No. FKZ  
\mbox{50 OR 0206} and \mbox{50 OR 0404}.
\end{acknowledgements}
 
\bibliographystyle{aa}
\bibliography{aamnemonic,myrefs}
 
\end{document}